\documentclass[%
 aip,
 amsmath,amssymb,
preprint,%
]{revtex4-2}

\usepackage{graphicx}% Include figure files
\usepackage{dcolumn}% Align table columns on decimal point
\usepackage{bm}% bold math
\usepackage{comment}
\usepackage{CJKutf8}

\usepackage[utf8]{inputenc}
\usepackage[T1]{fontenc}
\usepackage{mathptmx}
\usepackage{etoolbox}

\usepackage{setspace}

\makeatletter
\def\@email#1#2{%
 \endgroup
 \patchcmd{\titleblock@produce}
  {\frontmatter@RRAPformat}
  {\frontmatter@RRAPformat{\produce@RRAP{*#1\href{mailto:#2}{#2}}}\frontmatter@RRAPformat}
  {}{}
}%
\makeatother

\newcommand{\dd}{\mathrm{d}}

\begin{document}
\begin{CJK*}{UTF8}{gbsn}

\preprint{Accepted Manuscript}

\title{The Maintenance of Coherent Vortex Topology by Lagrangian Chaos in Drift-Rossby Wave Turbulence}% Force line breaks with \\
\thanks{This article may be downloaded for personal use only. Any other use requires prior permission of the author and AIP Publishing. This article appeared in \textit{Physics of Fluids} 36, 061701 (2024) and may be found at \url{https://doi.org/10.1063/5.0207687}}

\author{Norman M. Cao}
\altaffiliation{Corresponding author}
 \email{norman.cao@austin.utexas.edu}
\affiliation{%
 Institute for Fusion Studies, The University of Texas at Austin, 1 University Station, Austin, Texas 78712, USA
}%

\author{Di Qi (祁迪)}
 \email{qidi@purdue.edu}
\affiliation{
 Department of Mathematics, Purdue University, 150 North University Street, West Lafayette, Indiana 47907, USA
}%

\date{\today}% It is always \today, today,

%\begin{comment}
\begin{abstract}
This work introduces the ``potential vorticity (PV) bucket brigade'', a mechanism for explaining the resilience of vortex structures in magnetically confined fusion plasmas and geophysical flows.
Drawing parallels with zonal jet formation, we show how inhomogeneous patterns of mixing can reinforce, rather than destroy non-zonal flow structure.
We accomplish this through an exact stochastic Lagrangian representation of vorticity transport, together with a near-integrability property which relates coherent flow topology to fluid relabeling symmetries.
We demonstrate these ideas in the context of gradient-driven magnetized plasma turbulence, though the tools we develop here are model-agnostic and applicable beyond the system studied here.
\end{abstract}

\maketitle
%\tableofcontents
%\end{comment}

\end{CJK*}

Despite the conception of turbulence as a mixing phenomenon, large-scale coherent flows such as jets, vortices, and waves are found to coexist with turbulence in a broad range of natural and engineered systems.
Often associated with transport barriers known as Lagrangian coherent structures (LCSs),\cite{Haller2015} examples of particular interest to this work include Jupiter's alternating zonal bands and the coherent vortices within them,\cite{Marcus1993,Vasavada2005,Read2024} as well as Earth's jet streams and persistent weather patterns that arise from long-lived meanders known as atmospheric blocking.\cite{Wirth2018,Nakamura2018}
In magnetically confined fusion plasmas, important examples include the sheared \(E \times B\) flows that form in high-confinement-mode (H-mode) plasmas,\cite{Wagner2007} and zonally banded patterns of turbulence resembling ``\(E \times B\) staircases'' observed in simulations and experiments.\cite{Dif-Pradalier2015a,Hornung2017,Liu2021,Qi2021}

Due to the similarity in form between the magnetic and Coriolis force, strongly magnetized plasmas and strongly rotating geophysical flows exhibit strikingly analogous behaviors.
A key similarity is the conservation of potential vorticity (PV) in idealized models, a scalar material invariant linked to fluid element relabeling symmetries.\cite{Muller1995,Thorpe1995,Padhye1996}
Another key link is the presence of drift-Rossby waves, a class of unidirectionally propagating waves supported by fluid drifts in plasma contexts, and gradients of the Coriolis parameter or topography in geophysical contexts.

Understanding the dynamics of zonal jet flows and drift-Rossby waves in turbulent systems is crucial to unraveling the mechanisms of anomalous transport of heat and particles in magnetically confined fusion plasmas,\cite{Diamond2005} characterizing energy and momentum budgets in planetary climate systems,\cite{Wang2013,Woollings2023} and enhancing the predictability of extreme weather events.\cite{Altenhoff2008,White2022}
A major obstacle in this endeavour is the breaking of the statistical symmetries of turbulence by the large-scale structures.
An important example of this is given by the PV staircase, used to understand the resilience of zonal jets in planetary atmospheres and magnetized plasmas.\cite{Dritschel2008a,Gurcan2015}
In the staircase paradigm, zonal flows imprint onto the statistics of the turbulence, creating a spatially inhomogeneous patterning of mixing which in turn reinforces the existing zonal flow structure.
Spatial inhomogeneity precludes usage of the Wiener-Khinchin formula, so characterizations of turbulence in terms of Fourier power spectra cannot give a complete description of the two-point statistics of the velocity or other quantities, necessitating other approaches.\cite{Srinivasan2012,Parker2013}

\begin{figure}
    \centering
    \includegraphics[width=3.375in]{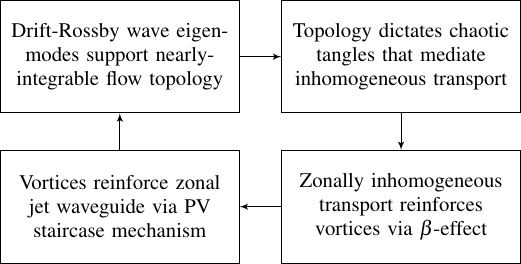}
    \caption{Flowchart illustrating key elements of the potential vorticity bucket brigade. The two boxes on the right describe the mechanism itself, while the two on the left describe its origins.}
    \label{fig:flowchart}
\end{figure}

In this work, we propose the ``potential vorticity bucket brigade'' as a way to understand the self-consistent transport effected by coherent vortices in 2d and quasi-2d drift-Rossby wave turbulence in the presence of zonal flows.
In analogy with the PV staircase, we argue that spatially inhomogeneous patterns of mixing due to the presence of coherent vortices leads to the reinforcement, rather than the destruction, of the vortices.
We demonstrate how this patterning arises from topological features of the Lagrangian flows associated with drift-Rossby eigenmodes in a zonal jet waveguide.
We approach this by addressing the four linked subproblems in Fig.~\ref{fig:flowchart}, which constitute the key elements of the mechanism.

\textit{Governing Equations}.---%
We study 2-dimensional plasma flows, and define unit vectors \(\hat{\mathbf{x}}\) pointing radially outward and \(\hat{\mathbf{y}}\) pointing zonally in the electron diamagnetic drift direction.
The geophysical equivalents are that \(\hat{\mathbf{x}}\) points northwards and \(\hat{\mathbf{y}}\) points westwards.
The background magnetic field points out of the plane in the direction \(\hat{\mathbf{x}} \wedge \hat{\mathbf{y}}\).
Rossby waves and electron-branch drift waves propagate in the `retrograde' direction \(\hat{\mathbf{y}}\) relative to the fluid medium.
`Co-rotating' vortices have an out-of-plane sense of rotation aligned with the magnetic field, and correspond to cyclones.

For concreteness, we develop the bucket brigade mechanism in the Dimits shift regime of flux-balanced Hasegawa-Wakatani (BHW) turbulence using simulation data from a previous study.\cite{Cao2023b}
The BHW model generalizes the Charney-Hasegawa-Mima (CHM) equations \cite{Connaughton2015} with a non-adiabatic electron response.\cite{Hasegawa1983,Majda2018}

We work in a doubly periodic domain \(D=[-L_x/2,L_x/2] \times [-L_y/2,L_y/2]\).
The model equations can be expressed in terms of the electrostatic potential \(\varphi(x,y,t)\), electron density fluctuation \(n(x,y,t)\), and (ion) vorticity \(\zeta(x,y,t)\) as
\begin{subequations}\label{eq:bhw}
\begin{gather}
    \label{eq:bhw_dens} \tilde{D}_t[n] := \left(\partial_t + \mathbf{u} \cdot \nabla - \mu \nabla^2\right)n = \alpha Q[\varphi - n] + \kappa u, \\
    \label{eq:bhw_vort} \tilde{D}_t[\zeta] = \tilde{D}_t[Q[n]] - \kappa u =: f, \\
    \zeta = \partial_x v - \partial_y u = \nabla^2 \varphi.
\end{gather}
\end{subequations}
Above, \(\mathbf{u} = (u,v) := (-\partial_y \varphi, \partial_x \varphi)\) is the \(E \times B\) velocity, analogous to the geostrophic flow.
\(\mu\) is both the viscosity and particle diffusivity, which are set equal here.
\(\tilde{D}_t\) is the combined advection-diffusion operator.
Length and time are normalized to the ion sound gyroradius \(\rho_s = 1\) and ion cyclotron frequency \(\omega_{ci} = 1\), equivalent to the Rossby deformation radius and Coriolis parameter respectively.

\(n\) is the fluctuation about some background density profile \(n_0(x)\) with fixed gradient \(n_0'(x) = -\kappa\).
Note \(n\) is analogous to layer height perturbations in geophysical contexts and \(\kappa\) is analogous to the \(\beta\)-effect parameter.
\(\alpha\) is the adiabaticity parameter, and models a non-adiabatic electron response due to finite resistivity.
\(Q[\chi] := \chi - \frac{1}{L_y}\int_{-L_y/2}^{L_y/2} \chi\, \mathrm{d}y\) is the non-zonal projection operator used to model the differing electron response to zonally-symmetric \(\varphi\).
\(f\) captures vorticity generation terms, which will be explained in detail later.
The BHW equations converge to the modified CHM equations in the adiabatic limit \(\alpha \to \infty\) and tend to have very strong zonal flows.\cite{Majda2018,Qi2019,Qi2020a,Qi2020b}

Turbulence in BHW results from a resistive drift-wave instability driven by the density gradient \(\kappa\).
The Dimits shift is a transitional regime where gradients are large enough to drive a primary linear instability but not large enough to develop a uniformly turbulent state.\cite{Dimits2000} 
This regime is expected to be relevant to magnetically confined fusion plasmas due to rapid increase of turbulent fluxes above marginality.\cite{Diamond1995}

Combining \(\eqref{eq:bhw_dens}\) and \(\eqref{eq:bhw_vort}\) gives the evolution equation for the BHW potential vorticity (PV) \(q + \kappa x\),
\begin{equation}
    \label{eq:bhw_pv}\tilde{D}_t[q+\kappa x] = 0
\end{equation}
where \(q := \zeta - Q[n]\).
In plasma contexts, the PV can also be related to the ion gyrocenter density.\cite{McDevitt2010,Madsen2015}
The velocity field cannot be recovered from the PV alone, so for studying flow patterns it is also useful to explicitly consider the transport and generation of the vorticity \(\zeta\).

\textit{Stochastic Lagrangian Representation}.---%
To link Eulerian and Lagrangian representations of the flow, we make use of the Feynman-Kac formula, which can be thought of as a stochastic generalization of the method of characteristics.

Fix some measurement time \(T\).
Following Ref.~\onlinecite{Eyink2020}, \(\tilde{D}_t\) is associated with the backwards It\^{o} stochastic differential equation (SDE) on \(s \le T\), 
\begin{subequations} \label{eq:sde}
    \begin{gather}
        \hat{\dd}\tilde{\mathbf{A}}^{s}_{T}(\mathbf{x}) = \mathbf{u}(\tilde{\mathbf{A}}^{s}_{T}(\mathbf{x}),s) \; \dd s + \sqrt{2 \mu} \; \hat{\dd}\tilde{\mathbf{W}}(s); \\
        \tilde{\mathbf{A}}^{T}_{T}(\mathbf{x}) = \mathbf{x}.
    \end{gather}
\end{subequations}
Here, \(\hat{\dd}\) denotes the backward It\^{o} differential and \(\tilde{\mathbf{W}}(s)\) is a vector Brownian motion.
\(\hat{\dd}\) is the time-reverse of the usual forward It\^{o} differential \(\dd\).
Reparameterizing time by \(\sigma := T - s\) would replace \(\hat{\dd}\) with \(\dd\) and \(\dd s\) with \(-\dd \sigma\).

The SDE \eqref{eq:sde} governs the motion of tracers starting from a given terminal point \(\mathbf{x} \in D\) at time \(T\) and flowing backwards in time parameterized by \(s\), perturbed by white noise.
\(\tilde{\mathbf{A}}^{s}_{T}(\mathbf{x})\) is a random variable taking values in \(D\) that tracks the tracer motion: given a realization of the noise \(\tilde{\mathbf{W}}(s)\), a tracer will flow from \((\mathbf{x},T)\) to \((\tilde{\mathbf{A}}^{s}_{T}(\mathbf{x}),s)\).
The stochastic backwards-time evolution arises naturally from considering the adjoint of the forward-time evolution associated with \(\tilde{D}_t\).

Often we want to integrate the vorticity \(\zeta(\mathbf{x},T)\) with respect to some function \(\eta(\mathbf{x})\).
Suppose \(\eta(\mathbf{x}) = w(\mathbf{x}) p(\mathbf{x})\) for some probability density \(p(\mathbf{x})\) and weight function \(w(\mathbf{x})\), and define a random variable \(\mathbf{X}\) distributed according to \(p(\mathbf{x})\).
Then for any time \(s < T\), via Feynman-Kac:
\begin{multline} \label{eq:feynmankac}
    \int_D \zeta(\mathbf{x},T) \eta(\mathbf{x}) \; \dd \mathbf{x}
    = \mathbb{E}\left[\zeta(\mathbf{X},T) w(\mathbf{X})\right] \\
    = \mathbb{E}\left[\zeta(\tilde{\mathbf{A}}^{s}_{T}(\mathbf{X}), s) w(\mathbf{X})
    + \int_{s}^{T} f(\tilde{\mathbf{A}}^{\tau}_{T}(\mathbf{X}), \tau) w(\mathbf{X}) \dd \tau \right]
\end{multline}
where expectations \(\mathbb{E}\) are taken over realizations of \(\tilde{\mathbf{A}}^{s}_{T}(\mathbf{X})\) with random terminal conditions \(\mathbf{X}\).

Physically, the first line expresses the Eulerian vorticity integral as a weighted average over a mass of infinitesmal Lagrangian fluid elements spread over \(p(\mathbf{x})\).
The second line relates the vorticity of these elements \(\zeta(\mathbf{X},T)\) to their vorticity at an earlier time \(\zeta(\tilde{\mathbf{A}}^{s}_{T}(\mathbf{X}), s)\) plus the net vorticity imparted over the interval \([s,T]\) by non-viscous forces \(f\).
The transport of vorticity by viscosity is accounted for by the stochastic motion.
Note that stochastic Lagrangian techniques can be generalized to include boundaries, 3d vortex stretching, nonlocal diffusion, and other effects.\cite{Eyink2009,Holm2015,Zhang2017,Eyink2020,Besse2023,Yang2023}

\textit{Vortex reinforcement by \(\beta\)-effect}.---%
One Eulerian characterization of vortices is that they correspond to peaks in the vorticity field \(\zeta\).
Here, we consider the coarse-grained vorticity \(\langle\zeta\rangle_\ell := g_\ell * \zeta\), which is convolved with \(g_{\ell}(\mathbf{x}) := \ell^{-2} g(\mathbf{x}/\ell)\) where \(g\) is a 2d unit Gaussian.

\begin{figure}
    \includegraphics[width=3.375in]{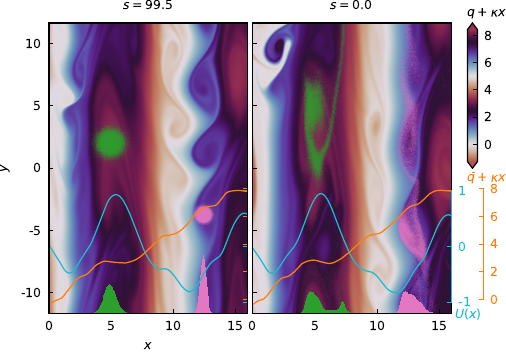}
    \caption{\label{fig:bhw_tracers}Two zoomed-in frames from a video showing stochastic tracers (green, pink) at different times \(s=99.5, 0\) plotted  on top of the potential vorticity field \(q+\kappa x\). Additionally, histograms of the \(x\) positions of the tracers are plotted at the bottom of the frames, showing the resulting skewed distribution of tracers. Profiles of the instantaneous zonally-averaged zonal flow \(U\) (cyan) and zonally-averaged PV \(\bar{q}+\kappa x\) (orange) are shown as well. (Multimedia available online)}
\end{figure}

We focus on two representative coherent vortices over a time interval \([0,99.5]\).
To compute \(\langle\zeta\rangle_{\ell}\) using \eqref{eq:feynmankac}, we fix a measurement time \(T=99.5\) and initialize a Gaussian patch of stochastic tracers at time \(s = T\) inside the vortices, then evolve them backwards in time until \(s = 0\).
This corresponds to taking \(\eta = g_\ell\) with \(w(\mathbf{x}) = 1\).
Representative frames from a video showing the evolution of the tracers are shown in Fig.~\ref{fig:bhw_tracers} (Multimedia available online).

Noting \eqref{eq:feynmankac} is linear in \(f\), we identify four contributions to the coarse-grained vorticity \(\langle\zeta\rangle_{\ell} = \langle\zeta_p\rangle_{\ell} + \langle\zeta_n\rangle_{\ell} + \langle\zeta_\kappa\rangle_{\ell} + \langle\zeta_\Delta\rangle_{\ell}\).
The first three are
\begin{subequations}
    \begin{align}
        \label{eq:z_coh} \langle\zeta_{p}\rangle_{\ell} &= \mathbb{E}\left[\zeta(\tilde{\mathbf{A}}^{s}_{T}(\mathbf{X}), s)\right], \\
        \label{eq:z_n} \langle\zeta_{n}\rangle_{\ell} &= \mathbb{E}\left[Q[n](\mathbf{X},T) - Q[n](\tilde{\mathbf{A}}^{s}_{T}(\mathbf{X}), s)\right], \\
        \label{eq:z_kappa} \langle\zeta_{\kappa}\rangle_{\ell} &= \mathbb{E}\left[\kappa \hat{\mathbf{x}} \cdot \left(\tilde{\mathbf{A}}^{s}_{T}(\mathbf{X}) - \mathbf{X}\right)\right].
    \end{align}
\end{subequations}
Meanwhile, \(\zeta_{\Delta}\) accounts for hyperviscous dissipation added for numerical stability as well as numerical error from discretizing the PDEs and SDEs, not explicitly included in \eqref{eq:feynmankac}.

\begin{figure}
    \includegraphics[width=3.375in]{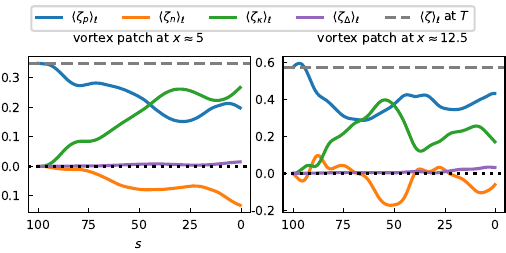}
    \caption{\label{fig:bhw_slbridge}Time traces of the contributions to the coarse-grained vorticity \(\langle\zeta\rangle_{\ell}\) for the two tracer patches in Fig.~\ref{fig:bhw_tracers}. \(s\) evolves backwards from the measurement time \(T\). The four solid lines in color will always add up to \(\langle\zeta\rangle_{\ell}\) measured at \(T\), shown in dashed gray.}
\end{figure}

The equivalence of the time integrals of \(f\) with the expressions above can be shown using It\^{o}'s lemma, and can also be seen as a consequence the PV evolution law (see supplementary material).
In Fig.~\ref{fig:bhw_slbridge}, we plot these quantities for the two patches of tracers shown in Fig.~\ref{fig:bhw_tracers}.
Note \(\langle\zeta_\Delta\rangle_{\ell}\) remains small, showing numerical error is negligible.

Now focusing on physical quantities, \(\zeta_{p}\) is the value of vorticity that would arise from passive stirring with no generation.
Due to turbulent mixing and dissipation, we expect \(\langle\zeta_{p}\rangle_{\ell}\) to decay to 0 as \(s\) gets further from \(T\).
If this process was dominated by the diffusive exchange of fluid, i.e. due to an impenetrable Lagrangian transport barrier surrounding the vortex, \(\langle\zeta_{p}\rangle_{\ell}\) would decay on the diffusive timescale \(\ell^2/\mu \sim O(10^{3})\).
Meanwhile, if it was dominated by turbulent mixing with no transport barrier, \(\langle\zeta_{p}\rangle_{\ell}\) would decay on the advective timescale \(\omega_{vortex}^{-1} \sim O(10^{1})\), estimated using the vortex rotation rate \(\omega_{vortex}\).
Fig.~\ref{fig:bhw_slbridge} shows that \(\langle\zeta_{p}\rangle_{\ell}\) decays at a rate between these two timescales, so neither limiting case applies.
This is supported by Fig.~\ref{fig:bhw_tracers}, which suggests a partial barrier to transport of tracers out of the vortex.

Now, we turn to \(\zeta_{n}\) and \(\zeta_{\kappa}\) to understand how vorticity generation balances turbulent losses.
These terms, corresponding to the two terms in \(f\), track changes in vorticity due to the accumulation of ion polarization charge necessary to maintain quasineutrality with the non-zonal \(Q[n]\) and background \(\kappa x\) components of the electron density respectively.
In geophysical contexts, \(\zeta_{n}\) and \(\zeta_{\kappa}\) would track changes in vorticity due to (non-topographic) vertical vortex stretching and the \(\beta\)-effect respectively.

Fig.~\ref{fig:bhw_slbridge} indicates that the \(\beta\)-effect term \(\langle\zeta_{\kappa}\rangle_{\ell}\) is responsible for maintaining the observed vorticity.
Since \(\langle\zeta_{\kappa}\rangle_{\ell}\) is proportional to the average change in the \(x\) position of the tracers, this process can be understood kinematically through the tendency for co-rotating vortices to entrain tracers from larger \(x\), as illustrated by the histograms in Fig.~\ref{fig:bhw_tracers}.
Similarly, counter-rotating vortices tend to entrain tracers from smaller \(x\).
Although this effect is well-known in geophysical contexts,\cite{Altenhoff2008,Pfahl2015} here we emphasize the analogy with PV staircases.
The coherent vortices are linked to partial transport barriers that create the zonally inhomogeneous mixing necessary to sustain the vortices at the observed amplitudes.

\begin{figure*}
    \includegraphics[width=7in]{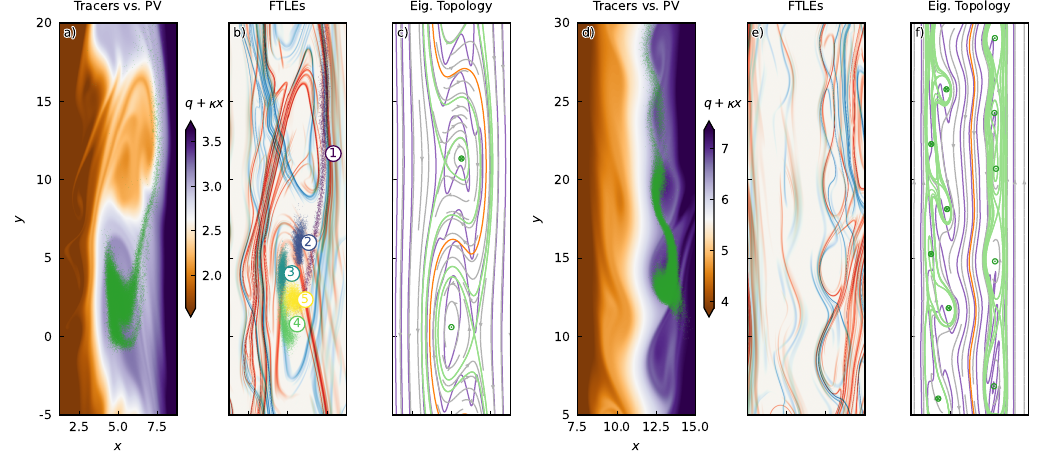}
    \caption{\label{fig:tangle} (a) Tracers in green overplotted on the PV field for one band of vortices, showing transversal filamentary structures. (b) Plot of forward and backward finite-time lyapunov exponents (FTLEs), with attracting/repelling exponents in red/blue respectively. The time history of a single tracer filament is overplotted. (c) Streamplot showing flows in the co-moving frame induced by drift-Rossby wave eigenmodes. Flow toplogy is shown with separatrices and O-points with sense of rotation in green and the nontwist torus in orange. Level sets of the wavy PV are shown in purple. (d-f) are similarly marked plots.}
\end{figure*}

\textit{Chaotic Tangles}.---%
To explain why this pattern of transport occurs, we first focus on the vortex patch located around \(x\approx 5\).
The distribution of tracers at \(s=68\) is shown in Fig.~\ref{fig:tangle}a.
Notice that the tracers exiting the vortex are drawn out into long filaments.
Close to where they attach to the vortex, the tracer filaments transversally intersect filaments in the PV field.

The transversally intersecting filaments relate to the presence of attracting and repelling Lagrangian coherent structures (LCSs) in the flow, collectively referred to as hyperbolic LCSs.
In planar flows, attracting (repelling) LCSs can be defined as material curves which have strong transverse attraction forward (backwards) in time.\cite{Haller2015}
By computing finite-time Lyapunov exponents (FTLEs), shown in Fig.~\ref{fig:tangle}b, we can heuristically detect the presence of hyperbolic LCSs.\cite{Haller2002}
Notice that PV filaments, advected forward in time, align with the attracting LCSs in red.
Tracer filaments, advected backwards in time, align with the repelling LCSs in blue.
Similar to past studies,\cite{Malhotra1998,Rogerson1999,Beron-Vera2010,Haller2012} we find transversally intersecting attracting and repelling LCSs, indicative of the chaotic tangle-like structure.

Using these LCSs, we can explain the tendency for chaotic transport to reinforce the vortex.
Fig.~\ref{fig:tangle}b shows the time evolution of one tracer filament over five equally spaced snapshots from \(s=0\) at step 1 to \(s=99.5\) at step 5.
In step 1, the tracer filament transversally crosses PV filaments outside the vortex.
Transport of PV into the filament is dominated by diffusion, akin to a `bucket' of fluid being filled diffusively with PV.

Backwards in time from step \(2 \to 1\), tracers are stretched by their attraction to the repelling LCSs.
Forwards in time \(1 \to 2\) this stretching is undone.
Advective transport then overtakes diffusive PV transport in steps \(2 \to 4\), akin to the `bucket' being passed along by the vortical flow.
Notably, in step \(3 \to 4\) the tracers pass through the vortex boundary in a manner reminiscent of lobes passing through a turnstile.\cite{Koh2000,Meiss2015}

Finally, in step \(5\) the tracers are fully entrained into the vortex, emptying the `PV bucket' into the vortex.
The net movement of the tracers was from larger to smaller \(x\), strengthening the vortex via the \(\beta\)-effect.
Many of these filaments form during the backwards evolution of the tracer patch, akin to a `brigade' of `PV buckets'.

Focusing now on the vortex patch around \(x\approx 12.5\), the filament kinematics are more complicated.
Comparing Fig.~\ref{fig:tangle}d to Fig.~\ref{fig:tangle}e, we can again see that tracer filaments tend to align with repelling LCSs and PV filaments align with attracting LCSs.
In contrast to the earlier case, tracers can leave the vortex through the transversal LCS intersections in either the `upstream' (prograde) or `downstream' (retrograde) directions.
Following the repelling LCSs, upstream tracers tend to be re-trapped into vortices leading to no overall transport in \(x\), whereas downstream tracers get lost at larger \(x\).
The net effect, as quantified in Fig.~\ref{fig:bhw_slbridge}, is that the vortex is reinforced.

\textit{Nearly-integrable Flow Topology}.---%
We now turn to explain how the tangled LCSs can arise from the self-consistent dynamics of the fluid.
Linearizing around the zonally and temporally averaged zonal flow and density profiles results in an eigenvalue equation for modes which propagate in \(y\) but are localized in \(x\) (see the supplementary material).
Plotted in Fig.~\ref{fig:tangle}c and f are approximations to the flow fields using the zonal flows plus 3-4 eigenmodes in each case.
These eigenmodes tend to have growth rates which are marginally unstable.\cite{Cao2023b}
Streamplots are shown for the instantaneous frozen velocity field in a frame of reference co-moving with the phase velocity of the dominant eigenmode.

Comparing Fig.~\ref{fig:tangle}c,f with Fig.~\ref{fig:tangle}b,e, there is a clear resemblance between the observed hyperbolic LCSs and the eigenmode streamline topologies.
Furthermore, these resemble topologies observed for Rossby waves propagating in the Bickley jet and the related standard nontwist map.\cite{DelCastilloNegrete1993,Del-Castillo-Negrete1996,Del-Castillo-Negrete2000}
The larger vortices exhibit a homoclinic nontwist topology, and are linked to drift-Rossby waves ``trapped'' in the retrograde jets.\cite{Zhu2020}
The smaller vortices exhibit a heteroclinic nontwist topology, and are linked to waves localized near the sharp PV interface associated with the prograde jets.

Despite their resemblance to linear structures, the vortices here are large amplitude.
PV perturbations from the vortices can be as large as the corrugations of the zonal PV, which in turn can be as steep as the background PV gradient.
Furthermore, for the large vortices the maximal vortex rotation rates \(\omega_{vortex} \approx 0.23\) are close to the maximal FTLEs \(\lambda_{lyap} \approx 0.23\), so the Kubo number \(Ku \sim \omega_{vortex} / \lambda_{lyap} \approx 1\) signals strong turbulence and the breakdown of quasilinear theory.\cite{Krommes2002,Diamond2010}
Note that trapping in elliptic (vortex-like) regions has been observed in other regimes of Hasegawa-Wakatani turbulence.\cite{Kadoch2022}

Linear dispersion is also not necessarily balanced against nonlinear effects, unlike the case for KdV-like drift-Rossby solitons.\cite{Redekopp1977}
For example, the variation in vortex sizes seen in in Fig.~\ref{fig:tangle}d evolves like an amplitude modulation of a \(\lambda_y = L_y/6\) carrier wave with an observed group velocity of \(v_{gr,obs} \approx -0.61\), close to the linear group velocity of \(v_{gr,eig} \approx -0.64\) and differing from the linear phase velocity of \(v_{ph,eig} \approx -0.4\).

To explain why the eigenmode flow topology persists, we build on the near-integrability property for drift-Rossby eigenmodes proposed in Ref.~\onlinecite{Cao2023a}.
The PV being an integral of motion in the ideal case implies a commutation condition between the Lagrangian flows and a fluid element relabeling symmetry generated by the PV, which can be used to extend Liouville integrability beyond Hamiltonian flows like the one studied here.\cite{Cao2023b}
Near-integrability states that the `wavy' eigenmode plus zonal background PV is a formally optimal near-integral of motion, suggesting that integrable flow topology due to relabeling symmetries can persist in wave-induced Lagrangian flows.
This is illustrated in Fig.~\ref{fig:tangle}c,f, where the contours of the wavy PV are nearly aligned with the eigenmode streamlines.
The alignment of the hyperbolic LCSs with the eigenmode streamlines shows that the formal near-integrability survives in practice, suggesting that these LCSs originate from separatrix splitting in the nearly-integrable flow.

\textit{PV Staircases}.---%
Finally, we observe that the vortices induce radially inhomogeneous mixing compatible with the PV staircase paradigm.
For the heteroclinic-type vortices in the prograde jets, mixing tends to occur on the flanks of the jets, further sharpening the PV jump which supports the jet.
For the homoclinic-type vortices in the retrograde jets, the non-twist torus has a much stronger meander.
Upon zonal averaging, this leads to the flattened PV region which supports the retrograde jets.
These jets then act as waveguides that localize the drift-Rossby eigenmodes, quenching the growth of most gradient-driven instabilities.\cite{Zhu2020,Cao2023b}
Only a small number of drift-Rossby wave eigenmodes are excited to large amplitude, playing the role of coherent horizontal convection rolls, and closing the loop on the bucket brigade.

\textit{Discussion}.---%
We summarize this work with two aphorisms covering the key elements in Fig.~\ref{fig:flowchart}.
\textit{``Topology begets flow'':} This work demonstrates that inhomogeneous mixing arising from flow topology is fundamental, rather than incidental, to the maintenance of flow patterns observed in a minimal model of gradient-driven drift-wave turbulence.
The coherent vortices support themselves through the entrainment of a `brigade' of 'PV buckets' whose kinematics are mediated by chaotic tangles in the Lagrangian flow.
\textit{``Flow begets topology'':} This topology is seen to arise from a near-integrability property of drift-Rossby wave eigenmodes arising from a fluid element relabeling symmetry.
Only a few of the eigenmodes are excited due to localization by the zonal jet waveguide maintained by radial transport from the vortices, closing the feedback cycle.

The PV bucket brigade mechanism is broadly consistent with universal pictures for spatiotemporal pattern formation in fluid systems near marginal stability.\cite{Crawford1991,Cross1993}
In this context, the exact stochastic Lagrangian representation of vorticity transport demonstrates that even when the eigenmode amplitudes are large, the net effect of the convective nonlinearity is to maintain the strength of the coherent flow patterns against mixing and dissipation.
Thus, the mechanism can be viewed as an argument for the applicability of single-mode descriptions of the flow.

Although the system studied here is highly idealized, the topological structures and other key parts of the bucket brigade mechanism align with phenomenology observed in similar physical systems.
The homoclinic-type vortices connected to drift-Rossby wave eigenmodes studied here bear strong resemblance to \(\Omega\)-type atmospheric blocks.\cite{Altenhoff2008,White2022}
The heteroclinic-type vortices resemble structures seen in Kelvin-Helmholtz wave instabilities \cite{Smyth2012} and high Reynolds number wall-driven 2d channel flow.\cite{Falkovich2018}
Near-integrability can also play a role in organizing flows with broad spectra of active modes.\cite{Cao2023a}
Given the rich variety of structures exhibiting Lagrangian coherence in nature,\cite{Haller2015,Rempel2023} it would be interesting to see if ideas based on stochastic Lagrangian techniques and relating integrability with relabeling symmetries can be used to describe structure formation beyond the system studied here.

%\begin{comment}
\section*{Supplementary Material}
A supplemental appendix that includes detailed equations, some additional derivations, and numerical methods is available online.

\begin{acknowledgments}
The authors thank P.J. Morrison and E. Vanden-Eijnden for valuable discussions that contributed to this work. N.M.C. acknowledges support by the US DOE under grant DE-FG02-04ER54742. D.Q. acknowledges support by ONR grant N00014-24-1-2192.
\end{acknowledgments}

\section*{Author Declarations}

\subsection*{Conflict of Interest}
The authors have no conflicts to disclose.

\subsection*{Author Contributions}
\textbf{Norman M. Cao:} conceptualization (equal), formal analysis (equal), methodology (equal), software (equal), writing -- original draft (equal).
\textbf{Di Qi:} conceptualization (equal), formal analysis (equal), methodology (equal), software (equal), writing -- original draft (equal).

\section*{Data Availability Statement}
The data that support the findings of this study are available from the corresponding author upon reasonable request.
%\end{comment}

\appendix

\section{Detailed Equations}

First, we derive the contributions to the coarse-grained vorticity in equations (5) in the main text by explicitly integrating the generation terms in the vorticity evolution equation.
Applying Feynman-Kac to equation (1b) in the main text, we  get
\begin{equation} \label{eq:char}
\begin{aligned}
    \zeta(\mathbf{x},T) &= \mathbb{E}[\zeta(\tilde{\mathbf{A}}_{T}^{s}(\mathbf{x}),s)] \\
    &+ \mathbb{E}\left[\int_{s}^{T} (\tilde{D}_t[Q[n]])(\tilde{\mathbf{A}}_{T}^{\tau}(\mathbf{x}),\tau) \; \dd{\tau}\right] \\
    &+ \mathbb{E}\left[\int_{s}^{T} -\kappa \mathbf{u}(\tilde{\mathbf{A}}_{T}^{\tau}(\mathbf{x}),\tau)\cdot\hat{\mathbf{x}} \; \dd{\tau}\right]
\end{aligned}
\end{equation}

Starting with the \(\zeta_\kappa\) term, note that the SDE (3) in the main text is equivalent to the stochastic integral
\begin{equation} \label{eq:pos_sint}
    \mathbf{x} - \tilde{\mathbf{A}}_{T}^{s}(\mathbf{x}) = \int_{s}^{T} \mathbf{u}(\tilde{\mathbf{A}}^{s}_{T}(\mathbf{x}),s) \; \dd{s} + \int_{s}^{T} \sqrt{2\mu} \hat{d}\tilde{\mathbf{W}}(s).
\end{equation}
The second term on the right-hand side of \eqref{eq:pos_sint} is a backward martingale which has zero average, so the expectation gives
\begin{equation}
    \mathbb{E}[\mathbf{x} - \tilde{\mathbf{A}}_{T}^{s}(\mathbf{x})] = \mathbb{E}\left[\int_{s}^{T} \mathbf{u}(\tilde{\mathbf{A}}^{s}_{T}(\mathbf{x}),s) \; \dd{s}\right].
\end{equation}
Taking the dot product of this equation with \(-\kappa \hat{\mathbf{x}}\) recovers the \(\zeta_{\kappa}\) term, and averaging over \(\mathbf{X}\) gives the expression for \(\langle\zeta_{\kappa}\rangle_{\ell}\) in the main text.

The integral for \(\zeta_n\) can be evaluated via It\^{o}'s lemma,
\begin{multline}
    \hat{\dd}Q[n](\tilde{\mathbf{A}}_{T}^{s}(\mathbf{x}),s) = (\tilde{D}_t[Q[n]])(\tilde{\mathbf{A}}_{T}^{s}(\mathbf{x}),s) \; \dd{s} \\ + \sqrt{2 \mu} \; \hat{\dd}\tilde{\mathbf{W}}(s) \cdot \nabla Q[n](\tilde{\mathbf{A}}_{T}^{s}(\mathbf{x}),s)
\end{multline}
In the stochastic integral form, the second term on the right-hand side corresponds to a backwards martingale which has zero average, so
\begin{multline}
    \mathbb{E}\left[Q[n](\mathbf{x},T) - Q[n](\tilde{\mathbf{A}}_{T}^{s}(\mathbf{x}),s)\right] \\
    = \mathbb{E}\left[\int_{s}^{T} (\tilde{D}_t[Q[n]])(\tilde{\mathbf{A}}_{T}^{\tau}(\mathbf{x}),\tau) \; \dd{\tau}\right]
\end{multline}
which recovers the \(\zeta_n\) term.

We remark here that BHW differs from the modified Hasegawa-Wakatani (MHW) equations in this term, as the vorticity generation term in MHW is \(f_{MHW} := \tilde{D}_t[n] - \kappa u\) compared to the `flux-balanced' \(f = \tilde{D}_t[Q[n]] - \kappa u\) used by BHW.
This change guarantees zero net particle flux in the adiabatic limit for BHW.

The contributions to the coarse-grained vorticity can also be derived from the the PV evolution law.
For brevity, let \(\mathcal{Q} := q + \kappa x\) be the BHW PV.
From It\^{o}'s lemma, we have
\begin{equation}
    \hat{\dd}[\mathcal{Q}(\tilde{\mathbf{A}}_{T}^{s}(\mathbf{x}),s)] = \hat{\dd}\tilde{\mathbf{W}}(s) \cdot \nabla \mathcal{Q}(\tilde{\mathbf{A}}_{T}^{s}(\mathbf{x}),s)
\end{equation}
where we have used the fact that \(\tilde{D}_t[\mathcal{Q}] = 0\), so the drift term is zero.
Taking the expectation of the integral form of the above equation gives
\begin{equation}
    \mathbb{E}[\mathcal{Q}(\tilde{\mathbf{A}}_{T}^{s}(\mathbf{x}),s)] = \mathbb{E}[\mathcal{Q}(\mathbf{x},T)]
\end{equation}
Substituting \(\mathcal{Q} = \zeta - Q[n] + \kappa x\) on both sides and rearranging gives the desired terms for \(\zeta_p\), \(\zeta_n\), and \(\zeta_{\kappa}\).

To derive equations for the eigenvalues and eigenfunctions, take a background zonal flow profile \(U(x)\) and split the fields into zonal and fluctuating components \(q = \bar{q}(x) + \hat{q}(x) e^{i (k_y y - \omega t)}\), \(n = \bar{n}(x) + \hat{n}(x) e^{i (k_y y - \omega t)}\).
Then,
\begin{widetext}
\begin{subequations}\label{eq:eig}
    \begin{gather}
        \omega \hat{q}(x) = k_y U(x) \hat{q}(x) - k_y (\bar{q}^{\prime}(x) + \kappa) \hat{\varphi}(x) + i\mu (\partial_x ^2 - k_y^2) \hat{q}(x) \label{eq:eig_q}, \\
	\omega \hat{n}(x) = k_y U(x) \hat{n}(x) - k_y (\bar{n}^{\prime}(x) - \kappa) \hat{\varphi}(x) + i\alpha(\hat{\varphi}(x) - \hat{n}(x)) + i \mu (\partial_x^2 - k_y^2) \hat{n}(x), \\
	\hat{\varphi}(x) = (\partial_x^2 - k_y^2)^{-1} \left[\hat{q}(x) + \hat{n}(x)\right].
    \end{gather}
\end{subequations}
\end{widetext}

The PDEs (1) in the main text are solved using a pseudo-spectral method with a 4th-order Runge-Kutta scheme for the time integration.
The SDEs (3) for the Feynman-Kac formula in the main text are solved using a 4th-order Runge-Kutta scheme for the deterministic part, and a Maruyama scheme for the stochastic part.
Grid-based data from the simulations are interpolated using 3rd-order Hermite polynomial interpolation in space and linear interpolation in time.
The expectation over terminal conditions and realizations of the Brownian noise is approximated by Monte Carlo sampling.
The eigenfunction equations \eqref{eq:eig} are discretized using a spectral method.

%\bibliography{main}% Produces the bibliography via BibTeX.

%aipnum4-2.bst 2019-01-14 (MD) hand-edited version of apsrev4-1.bst

\end{document}